\documentclass[preprint,12pt]{aastex}
\usepackage{natbib}
\begin{document}
\newcommand{\be}{\begin{equation}}
\newcommand{\bel}[1]{\begin{equation}\label{eq:#1}}
\newcommand{\ee}{\end{equation}}
\newcommand{\bd}{\begin{displaymath}} 
\newcommand{\ed}{\end{displaymath}}   
\newcommand{\bea}{\begin{eqnarray}}
\newcommand{\beal}[1]{\begin{eqnarray}\label{eq:#1}}
\newcommand{\eea}{\end{eqnarray}}
\newcommand{\e}[1]{\label{eq:#1}}
\newcommand{\eqref}[1]{\ref{eq:#1}}

\newcommand{\bfr}{{\bf r}}
\newcommand{\bfrp}{{\bf r'}}
\newcommand{\Scal}{{\cal S}}
\newcommand{\Jcal}{{\cal J}}
\newcommand{\Bcal}{{\cal B}}
\newcommand{\Jbar}{{\bar J}}
\newcommand{\Sbar}{{\bar S}}
\newcommand{\Jcalbar}{{\bar \Jcal}}
\newcommand{\Ibg}{{I_{\rm bg}}}
\newcommand{\etal}{et~al.}
\def\etal{{et~al.}}
\def\msun{$\rm M_\odot$}

\title{
Observations on the Formation of Massive Stars by Accretion}
\author{Eric Keto\altaffilmark{1} and
Kenneth Wood\altaffilmark{2},
}
\altaffiltext{1}{Harvard-Smithsonian Center for Astrophysics,
60 Garden Street, Cambridge MA 02138}
\altaffiltext{2}{University of St. Andrews, North Haugh, St Andrews, Fife,
Scotland, KY16 9SS}

\begin{abstract}
Observations of the H66$\alpha$ recombination line from the ionized gas in the
cluster of newly formed massive stars, G10.6--0.4, show that most of the
continuum emission derives from the dense gas in
an ionized accretion flow that forms an ionized disk or torus around 
a group of stars in the center of the cluster.
The inward motion observed in the accretion flow suggests that
despite the equivalent luminosity and ionizing radiation of several 
O stars, neither radiation
pressure nor thermal pressure has reversed the accretion flow. 
The observations indicate why
the radiation pressure of the stars and the thermal pressure of the HII
region are not effective
in reversing the accretion flow. The observed rate of the accretion flow, 
$10^{-3}$ M$_\odot$ yr$^{-1}$, is sufficient to form massive stars within
the time scale imposed by their short main sequence lifetimes.
A simple model of disk accretion relates quenched HII regions, trapped
hypercompact HII regions,  and photo-evaporating disks in an evolutionary
sequence

\end{abstract}

\section*{Introduction}

The formation of massive stars by the same accretion processes thought to form
low mass stars appears problematic because the intense radiation pressure from the
star and the thermal
pressure from the HII region around the star
may be sufficient to reverse the accretion flow and prevent matter from reaching the star.
Simple scaling arguments in purely spherical geometry first proposed decades ago
indicate that the luminosity to mass ratio of stars is
sufficient to limit the growth of stars by accretion to about 8 M$_\odot$.
How then does accretion proceed through the pressure of the intense radiation field to form
higher mass stars?
A second limitation on stellar mass is posed by the ionizing radiation
of early type stars. Stars of type B and earlier are hot enough to maintain
an HII region around the star. The temperature of this ionized gas is two
orders of magnitude greater than that of the surrounding molecular gas with a comparable
increase in pressure. How does accretion proceed against the outward
pressure of the HII region? Finally,
does the short main sequence lifetime of massive stars, a lifetime that decreases 
with increasing stellar mass, impose a third limitation?
For a star to gain the mass of an early O star before it evolves off the main sequence
may require a very high accretion rate,  $\dot {\rm M} > 10^{-3}$ M$_\odot$ yr$^{-1}$.
Since these questions were first posed,
more detailed theoretical investigations have found a number of ways for
accretion to proceed past these limits, and observational investigations have
found indirect evidence for the formation of massive stars by accretion, 
for example bipolar outflows and
disk-like structures around stars as early as type B. 
Nonetheless,
the question of how massive stars form by accretion remains.
While the several theoretical hypotheses appear sound enough, and the previous observational
evidence strongly implicates accretion, the observations have not so far demonstrated how
stars solve the problem of radiation and thermal pressure. Here we show for the first
time observations of the formation of massive stars with sufficient detail to
see how the accretion flow is structured to pass through the radiation and thermal
pressures. These new observations show that, exactly as suggested by previous theory,
non-spherical geometry and the high density and momentum of massive accretion flows
are the significant factors.

\section{Theoretical Backround}
\subsection{The Precursors to Massive Stars}

Early calculations of stellar evolution indicated that massive stars have 
very short pre-main sequence (PMS) lifetimes. For example, Iben (1965) 
estimated that
a 15 M$_\odot$ star has a PMS lifetime of $6\times 10^4$ yrs versus 
$5\times 10^7$ yrs for a 1.0 M$_\odot$ star. The short time scale 
suggests that unless a massive star 
forms with a very high accretion rate,
$\dot {\rm M} > {\rm M}_{star}/t_{PMS} \sim 2.5 \times 10^{-4}$ M$_\odot$ yr$^{-1}$, 
the star will
begin burning hydrogen while still in the accretion phase.  
Iben's calculations were for stars of a fixed mass and for a maximum
mass of 15 M$_\odot$. 
Stahler, Shu \& Taam (1980, 1981a, 1981b) and Palla and Stahler (1993) 
calculated the PMS evolution of stars 
growing by accretion and confirmed that stars
greater than several $M_\odot$ have no PMS phase.
\cite{BeechMitalas1994}, \cite{NorbergMaeder2000}, and \cite{BehrendMaeder2001},
extended these results for 
stars up to 60 M$_\odot$ for various accretion rates.

An example calculation (provided by Alessandro Chieffe based on the models
in Chieffi, Staniero, \& Salaris 1995) of the evolution of an accreting star 
starting from
an initial mass of 1.0 M$_\odot$ is
shown in figure \ref{fig:hr}. The evolutionary tracks for different accretion
rates intersect the main sequence
line at about 4 M$_\odot$ and follow that line as the star gains mass. These tracks
indicate that a star that is gaining mass by accretion
would appear little different in luminosity and temperature 
from a main sequence star of fixed mass.
Thus the stellar structure calculations suggest that if
massive stars form by accretion 
then the precursors to massive
stars would not resemble protostars in a PMS phase. Rather, the precursors to
massive stars would
appear similar to main sequence A or B stars except that because of the ongoing
accretion, they would likely have the characteristics associated with
accretion. For example, they would be deeply embedded, show evidence for
disks and bipolar outflows and possibly show maser emission from molecular gas
excited by the stellar radiation.

\subsection{Radiation Pressure}

The simplest
comparison between the outward force of the radiation pressure and
the inward force of gravitational attraction, 
$F_{rad}/F_{grav} = [\kappa L / (4 \pi r^2C) ] / (GM/r^2)$
indicates that stars of mass $>$ 8M$_\odot$ have sufficient luminosity
to counter the gravitational attraction driving accretion
(e.g.~Beech \& Mitalas 1994; 
Larson \& Starrfield 1971; Yorke \& Krugel 1977).
Here $\kappa$ is
the dust opacity, $L$ the luminosity, $r$ the radius that cancels out of the
equation, $C$ the speed of light, and $G$ the gravitational constant.
Although this would seem to imply that stars of mass greater than 8 M$_\odot$
could not form by accretion, the comparison is too naive. Kahn (1974) and
\cite{WolfireCassinelli1987} showed that if the dust is destroyed out to
some radius from the star, and if the accretion flow has sufficient momentum
to push the dust grains into the zone of destruction, the accretion flow will
be pushed into the dust free zone that is largely transparent to radiation.
\cite{Nakano1989} showed that 
the flattening of the accretion flow, as would be caused by rotation or
magnetic fields, would allow thermal re-radiation from the dust grains to
escape out the poles thus reducing the radiation pressure behind the
ionization front. The same reduction in pressure will occur
if the accretion flow is clumpy and the secondary radiation emitted by the dust
can escape between the clumps.
\cite{JijinaAdams1996} suggested that the radiation pressure in a 
rotating accretion flow would first cause the gas to fall onto the accretion disk
at larger radii rather than immediately reversing the flow as in the
case of spherical accretion. These several more detailed calculations suggest 
that radiation pressure might be a limiting
factor in high mass star formation, but the limiting mass should be larger
than 8 M$_\odot$.

\subsection{Thermal Pressure}

While the problem posed by the radiation pressure has received more attention
in the literature, 
the thermal pressure of ionized gas around the star
is also capable of reversing the
accretion flow
(Nakano, Hasagawa \& Norman 1995). 
All stars appear to form 
within dense clouds of molecular gas.
Stars of spectral type B and earlier  produce significant 
radiation shortward of the Lyman continuum limit that will ionize the molecular
gas and form an HII region around the star. 
Because the temperature of the
ionized gas is on the order of $10^4$ K while the temperature of the molecular gas 
is of order 100 K, the pressure of the ionized gas will be 100 times greater
than in the surrounding molecular gas, and the HII region should expand outward
reversing the accretion flow. This might imply that stars of spectral
type B would be unable to grow by accretion once an HII region formed around
the star. \cite{Keto2002b} suggested that the increased pressure of the ionized gas
would be irrelevant to the accretion flow as long as the zone of ionization
were within the sonic radius where the escape speed from the star exceeded the
sound speed of the ionized gas. This would occur  if the gas in the accretion
flow were dense enough that the radius of ionization equilibrium, where
recombination balances the ionization, were smaller than the sonic radius.

\subsection{The short main sequence lifetime}

Even without the problems posed by the radiation pressure and thermal
pressure,
the short main sequence lifetime of high mass stars compared with 
the accretion timescale might be the limiting factor in high mass star
formation
(Nakano 1989, Bernasconi \& Maeder 1996, Behrend \& Maeder 2001).
If the accretion rate is not high enough to supply sufficient fresh hydrogen
to dilute the helium ash of hydrogen burning, then the star will evolve off the
main sequence where stellar pulsations and the eventual ejection of the star's envelope
will disrupt and end the accretion onto the star. Figure \ref{fig:hr} shows that the accretion
rate has to be quite high $\dot M > 10^{-3}$ M$_\odot$ yr$^{-1}$ to keep a
growing star on the main sequence long enough to reach the mass of an O star.
Because the main sequence lifetime is a decreasing function of stellar mass
as indicated in figure 1,
the accretion rate might have to increase as
some function of the mass of the star.  
There is observational evidence that the accretion rate depends on the stellar mass
from observations of both high and low mass stars (Churchwell 1998; Henning et al.~2000;
Calvet et al.~2004).

\subsection{Observational indications of accretion}

A number of previous observations provide indirect evidence for the formation of
high mass stars by accretion. Accretion flows toward massive stars have been observed 
in several high
mass star forming regions (Ho \& Haschick 1986; Keto, Ho, \& Haschick 1987, 1988; 
Young, Keto, \& Ho 1998; Zhang et al.~1998, 2002; Kumar et al.~2002, 2003).
Massive accretion disks around high mass stars
have been detected by both imaging and spectroscopic means
(Zhang et al.~2002; 
Sandell et al.~2003; Olmi et al.~2003; Beltran et al.~2004; Chini et al.~2004). 
Evidence for bipolar outflows too
massive to be powered by low mass stars also  indicates massive accretion flows
(Shepherd \& Churchwell 1996; Beuther et al.~2002). 
Observations of stellar photospheres and bipolar outflows also indicate that the accretion rates 
depend on the stellar mass (Churchwell 1998; Henning et al.~2000; Calvet et al.~2004), 
as suggested by the theory of continuing accretion.
Radio frequency spectral 
energy distributions observed from hypercompact HII regions
are consistent with ionized accretion flows onto massive stars 
as pointed out by van der Tak \& Menten (2005) and inferred retrospectively in
the observations of  
Hofner et al.~(1996) and  Testi et al.~(2000)

However, 
observational evidence that directly demonstrates the
validity of the several theoretical hypothesis as to exactly how
high mass stars form by accretion has been lacking. 
There are three principle difficulties in observing the formation of high
mass stars: distance, extinction, and the short time scale for formation.
Because most of the massive stars, with a few exceptions such as the Orion
star forming region, are at kpc distances, it is difficult to obtain the
sensitivity and angular resolution to measure the properties of accretion flows
around individual high mass stars. Secondly, massive stars form within
very dense molecular clouds with enormous extinctions through the near
infra-red and even the millimeter radio. Third, because of the exceptionally 
short time scales, not many examples
of massive O stars in formation would be expected.

These observational difficulties can be overcome.
Centimeter wavelength radio waves are not much affected by dust emission and 
absorption and can be used to see into the densest molecular clouds.
Even at large distances, 
accretion flows
onto small groups of stars, with a higher combined mass, $M$, than single stars,
have accretion flows with velocities, typically proportional to $\sqrt{GM}$, 
and size scales, proportional to $GM/c^2$, for sound speed $c$, that
are large enough to resolve with radio interferometers.

There is a second reason to target accretion flows onto small groups of massive
stars. Formation in groups of a few stars may be the common mode
in high mass star formation.
The scalings above suggest that if massive stars form by accretion and the
accretion rate depends on the stellar mass, then massive stars are 
more likely to form in binaries or small groups that can generate the massive
accretion
flows needed to overcome radiation and thermal pressure and
the short main sequence time scale.
There is also indirect observational
evidence for this proposition.
For example, observations by 
Okamoto et al.~(2003) show that a number of HII regions contain several stars each.
Prebisch et al.~(2001) and Mermilliod \& Garcia (2001)
observed that massive stars form in
binaries and small groups of three or four at a greater frequency than do low mass stars.

Although accretion flows onto groups of stars may be different than onto single stars, these
flows may be characteristic of the accretion flows in high mass star
formation.
Thus we chose as a target for our observations the accretion flow onto a small group
of stars at the center of the 
G10.6--0.4 cluster 
(Ho \& Haschick 1986, Keto 2002a, Sollins et al. 2005).

\section{Observations}
The new observations were made with the Very Large Array (VLA)
\footnote{The National
Radio Astronomy Observatory is
a facility of the National Science Foundation operated under
cooperative agreement by Associated Universities, Inc.}
radio interferometer
on February 23, 2002.  The array was in the "A" configuration with baselines
up to 36 km leading to an angular resolution of $0.2 \times 0.1$ arcsec.
The correlator was configured for a bandpass of 12.5 MHz with 64 channels of
195.31 kHz or 2.62 kms$^{-1}$ at the observing frequency of the H66$\alpha$
recombination line of hydrogen at 22.36416932 GHz. 
The data were
combined with H66$\alpha$ observations made in the D configuration and 
described separately in Keto (2002a) and images and spectra made with
the combined data set.
The sum of the channels off
the spectral line yields a map of the continuum emission that when subtracted
from the total emission yields a map of the H66$\alpha$ spectra. 
The $1\sigma$ noise level in the data is 0.7 mJy/beam per channel.
After subtraction of the continuum flux, the noise level in the maps
of line emission was 1.2 mJy/beam per channel.

\subsection{The Accretion Flow}

Figures \ref{fig:continuum} and \ref{fig:arcs} 
show the continuum emission around the group of stars. The
plane of the accretion disk is indicated
in figure \ref{fig:continuum}
by the solid line at 45$^\circ$ across the figure. The orientation of 
a high velocity outflow perpendicular to the disk is shown by the dotted line.
Figure \ref{fig:arcs} is a detail of photo-ionized arcs, previously identified
in Sollins et al.~(2005) to the left of the bright HII region.
The location is marked as "ARCS" in figure \ref{fig:continuum}.

The bright continuum emission seen in the new high angular 
resolution observations of G10.6--0.4 is
identified as a dense accretion disk or torus rather than a classical ionization bounded HII
region by the orientation of the emission with respect to the accretion
flow and by the presence of the nearby photo-ionized arcs.  Previous observations 
(Keto 1990, Sollins et al. 2005) of
molecular emission and absorption in G10.6--0.4 have mapped the structure
of the molecular accretion flow onto the stars and identified the
axis of rotation as projected onto a northeast-southwest direction
(dashed line, figure 2).
If the continuum emission were from a classical HII region with a boundary defined by
equilibrium between ionization and recombination, the HII region would be
extended down the density gradient in the direction of the lower density gas,
perpendicular to the rotational flattening of the accretion flow.
In this case the continuum emission would
have an elliptical shape elongated in the direction of the rotation axis.
However, 
the observations, figure \ref{fig:continuum}, show that the continuum emission is
seen elongated along the equatorial plane 
following the densest gas of the accretion flow.

The absence of an ionization boundary around the bright continuum emission is
also indicated by the photo-ionized arcs 
most prominently on the southeast (left)
side of the central ellipse, figures \ref{fig:continuum} and \ref{fig:arcs}. 
The orientation of the photo-ionized arcs
with their ionization fronts facing the stars 
indicates that these outlying clumps of high density molecular gas
receive ionizing radiation from the stars. This would not be possible
if there were an ionization front at the  boundary of the bright continuum emission.
The gas between the bright continuum ellipse
and the photo-ionized arcs must be ionized to allow the stellar radiation to pass
and reach the arcs, and the low continuum brightness
between the arcs and the disk
confirms that this gas has a lower density.
Thus the apparent boundary of the
bright continuum ellipse is set by the outwardly decreasing density of the ionized gas
and the reduction in emitted continuum flux as the emission measure, $n_e^2L$,
rather than by an ionization boundary.

While
the bright continuum emission indicates the morphology of the densest
ionized gas around the group of stars,
the H66$\alpha$ line provides the information on the velocities within the ionized gas
to identify this dense gas as an accretion disk or torus. 
In our data we are also able to identify two other components of the accretion flow, infall in a
quasi-spherical envelope and a high velocity outflow
aligned with the rotation axis. 

Figure 4 shows
the trend of velocity with position in the direction along the equatorial
plane of the disk, at the position marked by the solid line in figure \ref{fig:pv_LONG_disk}.
The backward "C" of emission seen in 
figure \ref{fig:pv_LONG_disk} is characteristic of radial or spherical infall.
The "C" structure arises because
on  lines of sight progressively closer toward the center of the infall,
the velocity of the emission is more red-shifted since the velocity of the
radial infall is projected more directly along the line of sight.
The "C" shape is characteristic of emission from the front hemisphere of a spherical
inflow in which the velocities are directed radially inward. 
Emission from the back hemisphere would produce a corresponding "C" in the
opposite orientation. Optical depth effects are likely responsible for the
weakness of the emission from the back of the flow (Keto 2002a).
The velocity of the quasi-spherical infall in the ionized envelope is
indicated at $> 10 $ kms$^{-1}$ and the rotation at $> 10 $ kms$^{-1}$.

More rapid rotation of the accretion disk at a smaller scale toward the
center of the group of stars is indicated by the breadth of the  emission lines in
the center of figure \ref{fig:pv_LONG_disk} and by the linear gradient in line center
velocities across the figure, along the plane of the disk.  
The more rapid rotation toward the center of the disk and the
more modest rotation in the envelope together show that the accretion flow is spinning up
as the gas spirals in toward the stars. 

Figure \ref{fig:pv_SHORT_disk}
is a position-velocity diagram of the
H66$\alpha$ emission along the axis of rotation through the middle of the ionized disk
at the position marked by the dotted line in figure \ref{fig:continuum}.
A high velocity outflow of ionized gas is seen in figure \ref{fig:pv_SHORT_disk}
in the fainter emission
extending northeast from the center of the disk, from a V$_{\rm LSR}$ of near
zero kms$^{-1}$ to over 60 kms$^{-1}$ where
the recombination line emission falls below the sensitivity limit.
Referring back to the map of continuum emission,
figure \ref{fig:continuum}, the northeast outflow appears bounded by two lines of
continuum emission in a "V" shape, previously identified in Sollins et al.~(2005). 
There is a corresponding "V", of
wider opening angle, extending to the southwest indicating a counter outflow.
The opening angles of
both outflows are wider than those of high velocity jets associated with
individual high mass stars, and the outflow may be
driven by the collective effects, stellar winds or individual jets, of
the several massive stars in the central group.
The gas density inside the outflow is quite low as indicated by the weakness
of the continuum emission at the location of the outflow. 

The lines in the high velocity
outflows are narrower than in the center of the flow
consistent with lower density gas in the outflow and a smaller
velocity gradient. 
Figures \ref{fig:spectrum_center} through
\ref{fig:spectrum_southwest} show the spectra of the H66$\alpha$ recombination line
at three positions: the center of the flow at the intersection of the solid
and dashed lines in figure \ref{fig:continuum} and at two  positions 1$^{\prime\prime}$ away 
from this intersection along the dashed line (rotation axis) to the northeast 
(figure \ref{fig:spectrum_northeast}) and southwest (figure \ref{fig:spectrum_southwest}).
The breadth of the line $\sim 50$ kms$^{-1}$ in the center of the disk is consistent with
a combination of pressure broadening due to the
high density of the gas in center of the disk (Keto et al. 1995)
and  high rotational velocities
that are spatially unresolved.

\section{A simple model of the accretion flow}

The accretion flow onto G10.6--0.4 is quasi-spherical at larger distances ($> 0.1$ pc)
from the central group of stars where the flow is molecular and seen in absorption 
against the HII region (Keto 1990; Sollins et al.~2005). On the much smaller
scale of a few thousand AU where the flow is ionized and seen in emission, the 
flow begins to flatten into an accretion disk. 
The extent of the bright ionized emission in the plane of the accretion disk is
about 2$^{\prime\prime}$ indicating a radius of about 5000 AU using the
distance of 4.8 kpc derived by Fish et al.~(2003). If G10.6--0.4 is at the further
distance of 6 kpc derived by Downes et al.~(1980), the scale would be a 
little larger and the derived mass below would be correspondingly greater. 
The infall and rotational velocities in the overlying molecular flow are similar at
about 4.5 kms$^{-1}$ and 2.5 kms$^{-1}$ respectively (Keto 2002a), and the 
gas with these velocities must 
be outside of 5000 AU. The observed velocities in the ionized accretion flow
are also similar and both $>  10$ kms$^{-1}$, and the gas with these velocities
must be inside of 5000 AU. The comparison of the molecular and ionized 
velocities indicates that  the accretion flow spins up
as it approaches the stars. 

There is considerable uncertainty in pairing the gas velocity with a radius because
the observations average the emission along the line sight through the source.
Nevertheless,
the separation of the flow into molecular and ionized regions provides a definite
radius to work with.
Second, there is uncertainty in the velocities themselves which are likely to be lower limits. 
The two velocity components are both derived from the difference of the velocity of
the spectral lines line
toward the center of the HII region or disk and toward the edge, and the blending 
along the line of sight reduces this difference. For example, in the case of the inflow
velocity, because
the flow is accelerating, the maximum velocity at the center of the HII region is blended with
lower velocities from the overlying gas. Similarly, we never see the true tangential
minimum velocity at the edge of the HII region because of blending with higher 
velocity overlying gas.  Pressing on we may estimate the enclosed mass required to
generate these velocities as $M \sim Rv_T^2/G$
where the total velocity is the quadrature sum of the components, 
$v_T = \sqrt{v_{infall}^2 + v_{orbital}^2}$. Calculating the mass using a radius of
5000 AU and the two velocity components of the ionized gas at 10 kms$^{-1}$ 
each yields 1100 M$_\odot$.
Substituting the lower molecular velocities and assuming the same radius yields
a mass of about 200 M$_\odot$. The true enclosed mass may be between these
limits. For definiteness we will suppose 300 M$_\odot$ although the true mass
could be higher. For reasonable gas
densities, most of this mass must be in stars, and we assume that the mass is
divided among some number, perhaps  five, O stars.

A particularly simple model for an accretion flow that appears quasi-spherical
at large distances and flattens to a disk at smaller radii 
is that of the gas orbiting
on ballistic trajectories around a point mass (Ulrich 1976, Terebey, Shu \& Cassen 1984).
In this model for steady state
accretion, the self-gravity and pressure of the gas are assumed
negligible, the point mass at the
center of the flow is constant, and the cloud has a specific
angular momentum at infinity, $\Gamma_\infty$, set equal to solid body rotation, which
is then conserved on the flow trajectories.  Where the trajectories
would cross the mid-plane of the flow, the gas is assumed to collide
with a similar flow from the
other side of the plane so that the gas settles into a disk on the mid-plane.
The model produces a reasonable approximation within the gravitational
radius of the flow, $r_G = GM/c^2$,
where the velocities are supersonic and the thermal pressure of the gas is
negligible and
away from the mid-plane where the densities in the model are formally infinite.
The radius of disk formation, $r_D$, where the centrifugal force equals the gravitational
force, $\Gamma^2/r_D^3 = GM/r^2_D$,  divides the flow into an
outer region where the inflow is quasi-spherical and an inner region where
the flow is rotationally dominated.
Figure \ref{fig:streamlines_3d} shows several trajectories of such a flow with the radius of disk
formation, $r_D$, in non-dimensional units at $r_D = 1$. The density in the flow, just off the
mid-plane, is indicated
by the color map at the base of the figure.

The observations of the accretion flow in G10.6--0.4 relate to this model as follows.
The radiation of the stars is able to ionize the flow out to some distance. 
If the ionizing flux, the enclosed mass, and the density of the flow are
such that the radius of ionization is within the sonic radius where the escape speed
from the stars equals the sound speed, $c$, of the ionized gas, $r_{ionized} < GM/c^2$,
then the pressure of the ionized gas will be irrelevant to the dynamics of the flow
(Keto 2002b).
Thus although
the molecular flow will become ionized at some radius $r_{ionized}$, 
the increase in gas pressure due
to the ionization will not significantly counter the gravitational attraction of the stars, and
the accretion flow will transition from molecular to ionized while maintaining its
continuity. In G10.6--0.4 because the molecular accretion flow is clumpy and does not
have a smoothly increasing density gradient, the radius of ionization will be ragged
and defined by the photo-ionized edges of clumps interspersed in lower density fully
ionized gas. These edges appear as the photo-ionized arcs in figures 2 and 3.
Where the molecular gas is most clumpy, the variations in the radius of ionization will
be greatest. What is important for continuing accretion is that the inward flow,
which may consist entirely of molecular clumps within low density ionized gas, must
remain molecular until within the gravitational radius, $r_g$, of the ionized gas.
Despite the complexity imposed by the non-uniform density of the clumpy interstellar
medium, we will continue the
description of the model  assuming a smooth density profile. 

The simple scalings suggest that if the infall velocity is greater than the sound
speed of the ionized gas, $v_{infall} > c$, the HII region, or more generally, the
zone of ionization will be trapped within the accretion flow. Such trapping requires that the
average (smooth) density gradient should decrease outward no faster than
$n \sim r^{-3/2}$ (Franco, Tenorio-Tagle, Bodenheimer 1990). This condition is
generally satisfied by steady-state 
accretion models because the maximum infall velocity in steady
state accretion is the free-fall velocity, $v \sim r^{-1/2}$. Conservation of mass then
imposes a maximum limit on the density gradient of
$n \sim r^{-3/2}$. 
Of course, if the density gradient is set by conditions other than the accretion flow,
for example by the edge of the molecular cloud or the assumption of a particular density
profile, for example that of a singular isothermal sphere, then the density gradient may 
be steeper, and in that case  trapping is not possible. Notwithstanding, within the 
definition of the model of steady state accretion,
if the radius of ionization is within the gravitational
radius then trapping is inevitable.
 
As the accretion flow approaches the central mass, it will spin up and form a disk
at the radius, $r_D$ defined above. Since $r_D  \sim GM/v_{orbital}^2$, then if
$v_{orbital} < c$ at the gravitational radius, $r_G$, then the radius of disk formation 
will  be within the
maximum radius, $r_G$, of a trapped HII region. Thus the radius of disk formation,
depending on the flux of ionizing photons, could
be within the ionized portion of the flow meaning that the accretion disk would be formed
out of infalling ionized gas. Of course the accretion disk itself need not be fully
ionized. Within the definitions of the model, the density of the 
accretion disk becomes
arbitrarily high near the mid-plane, and therefore the center of the accretion disk may be neutral
if the rate of recombination, which scales as the density squared, exceeds the
rate of ionization. Nonetheless, the disk will be formed out of inward 
flowing ionized gas entirely within the ionized region of accretion flow.

In G10.6--0.4, the measured velocities  within the ionized flow are $\geq 10$ kms$^{-1}$,
comparable to the sound speed of the ionized gas. Given the uncertainties in
the velocities, it is possible that the conditions above that would result in a trapped
HII region, $v_{infall}(r_G) > c$, and an ionized accretion disk, $v_{orbital}(r_G)< c $,
would apply. The observed morphology of the accretion flow with a quasi-spherical molecular flow
at larger scales and an ionized accretion disk at smaller scales suggests
that these conditions apply.   

\section{A numerical model for the ionization structure of the accretion flow}

To illustrate the comparison between the theoretical and observed morphologies we
show the results of a calculation of the structure of an accretion flow subject to
a central source of ionizing radiation.
Figure \ref{fig:datamodel}  shows the estimated 1.3 cm free-free emission 
from a model accretion flow with  
a TSC density profile with an accretion rate of $10^{-4}$ M$_\odot$ discretized onto 
a $65^3$ linear Cartesian grid
$40\, 000$~AU on a side that is
subject to an ionizing luminosity of $2\times 10^{50}$ s$^{-1}$ in the continuum and
a spectrum equivalent to a $40\,000$~K, WM-basic
model atmosphere from the library of Sternberg et al. (2003).
(Because a TSC accretion flow is flattened due to rotation, the flow is concentrated
in a smaller solid angle than a spherical flow. Thus a TSC flow has a lower 
accretion rate than a spherical flow with the same inward momentum, $\rho v$.)
The ionization fraction and temperature of this model are
calculated with a 3D Monte Carlo photoionization code \citep{WoodMathisErcolano2004},
and the observable free-free emission follows from the
the free-free emissivity and opacity (Osterbrock 1989).

To further the comparison with the observations which show a clumpy rather
than smooth density profile in the accretion flow, and
to explore the effects of a clumpy accretion flow, we arbitrarily
place around the flow six spherical high density clumps with radii 
of 2000~AU and uniform densities, $10^5\,{\rm cm}^{-3}$
to suggest the photoionized arcs in our observation, figure  \ref{fig:arcs}.
An evacuated bipolar cavity is also included in the model as an approximation to
the low density bipolar outflow seen in the observations. 
We expect that an outflow would appear along the rotation axis
where the density gradient is steepest and  where 
any excess pressure will find the easiest direction of breakout.
But because
the outflow must be due to forces not
included in the infall model such as stellar winds or bipolar outflows from the
individual stars, the outflow in our model is simply arbitarily imposed on the smooth
TSC accretion flow.

The calculation shows that the gas in the accretion flow
is maintained in a highly ionized state, with the exception of the densest
regions deep within the disk or torus where the recombination rate exceeds
the rate of ionization.  Figure \ref{fig:ionstruc} shows the density,
and ionization structure and temperature on a one pixel wide slice through the model.
Also shown is the column density of neutral gas.
The model does not have the ionization front of a classical HII
region, but because the free-free emission scales as the square of
the density which falls off roughly as, but less steeply than,  $r^{-3/2}$, the emission appears 
bounded. 
In the model there is continuum flux from the low density ionized gas surrounding the
bright torus, but its level is below the detection limit of our observations.
However, as discussed above, the differences between an
accretion flow and a classical HII region can be discerned from the
observations.

\section{An estimate of the effect of radiation pressure on the accretion flow}

Once we have described the accretion flow and have
an estimate of the mass accretion rate and the inward velocity, 
we may compare the force deriving from the inward momentum of
the accretion flow with the outward force deriving from the radiation pressure.
The luminosity
estimated from infrared emission is
$1.2\times 10^6$ L$_\odot$ 
(Fazio  et al. 1978).
The most naive estimate of the force deriving from this luminosity, assuming 
spherical geometry and the total absorption of the luminosity by the dust in the flow, 
would be $L/c \sim 1\times 10^{29}$ dynes. 
A similarly simple estimate of the force deriving from the momentum of the
accretion flow, 
assuming 
a spherical accretion rate  of
$10^{-3}$ M$_\odot$ yr$^{-1}$ and velocity of 4.5 km s$^{-1}$ at 5000 AU (Keto 2002b),
would be,
$\dot {\rm M} v
\sim 3\times 10^{28}$ dynes.
The similarity of the two estimates
suggests that the inward momentum of the accretion flow
is of the magnitude required to compete with the radiation pressure even in
the absence of the mitigating effects of non-spherical geometry and a clumpy
flow that are both indicated by the observed structure. Thus the observations of
the accretion flow around G10.6--0.4 
suggest that accretion may proceed despite the high luminosity $> 10^6$ L$_\odot$
of the central stars.

\section{The relation between photoionized accretion disks in trapped hypercompact HII regions,
massive molecular accretion disks, and photoevaporating accretion disks}

The observations of G10.6--0.4 suggest the model of disk formation out of fully ionized
gas within a trapped HII region. 
The structure of this model  accretion flow depends on the relative magnitudes of
a few parameters. In particular, the extent of the ionization depends on the 
radiative flux of the stars and the
density of the surrounding gas. The dynamics of the accretion flow depend 
primarily on the mass of  the stars and the initial angular momentum of the surrounding
gas. Because the extent of the ionization and the dynamics of the accretion  depend
on different parameters, it is possible to imagine that  different
combinations of parameters will result in accretion flows with different structures.

For example, the structure in G10.6--0.4 with an accretion disk formed out of ionized
gas within the trapped HII region requires that $v_{orbital}(r_G) < c$ or more
specifically that $r_D < r_{ionized}$. Suppose that the initial conditions that led
to this flow instead included a higher angular momentum or lower ionizing flux  
so that $r_D > r_{ionized}$. In this case, our model accretion
flow would form a disk in the molecular portion of the flow.  
If in addition, the extent of the ionization
were within the sonic radius, $r_{ionized} < r_G$, there would be a small trapped
HII region at the center of the flow. If the HII region were small and dense, as would
be implied by its trapped condition, and perhaps
optically thick, the observed continuum flux would be very low.
This may be the appropriate model for those sources where a massive molecular 
accretion disk or torus is inferred from observations of  a high mass proto-stellar
candidate with no observed continuum emission or weak continuum emission
from a hypercompact HII region 
(Zhang et al.~2002; 
Sandell et al.~2003; Olmi et al.~2003; Beltran et al.~2004; Chini et al.~2004).

We may consider the evolution of this alternate model accretion flow with a 
massive molecular rather than ionized disk. In Keto (2003) the following
evolutionary scenario was proposed for trapped HII regions in a purely
spherical accretion flow. We will review this spherical model and then
apply the concepts to a model accretion flow with rotation and a disk.

The evolution in the spherical model takes place in three stages and the
accretion flow passes through the three stages as the central star or stars
gain mass and their temperatures and therefore their ionizing  fluxes increase with
their increasing masses. In the first stage when the ionizing flux is very low, the
HII region may be quenched (non-existent) if the flux of neutral hydrogen
atoms or molecules exceeds the flux of ionizing photons. In this case the HII
region will not extend beyond the boundary of the star. This is the model
discussed in Walmsley (1995). As the star or stars gain mass as a result of the
continuing accretion, the flux of ionizing photons will increase eventually
exceeding the flux of neutral hydrogen. At this point, provided that the
density gradient of the accretion flow does not exceed the limit inherent in
models of steady state accretion, $n\sim r^{-3/2}$, a trapped HII region will
develop around the star. This is the second stage of evolution.
As long as the limiting density gradient is
not exceeded, 
trapping is inevitable in this model and requires 
no fine tuning of the parameters. 
As the star or stars continue to gain mass by accreting matter through the
trapped hypercompact HII region, the temperatures and ionizing fluxes of 
the stars will further increase.
Correspondingly, the size of the trapped HII will grow, not by pressure driven expansion,
but simply owing to the increase in the ionizing flux. At some point, the size
of the HII region will exceed the sonic radius where the escape velocity
equals the sound speed of the ionized gas. In Keto (2002b) it was shown that
at this point, the HII will begin hydrodynamic expansion and transition rapidly to the
classical model described by pressure driven expansion (Spitzer 1978; Dyson \& Williams 1980;
Shu 1992). This is the third and final stage of evolution.

Let us consider what would happen in this evolutionary scenario
if the accretion flow included high rotation and a massive molecular disk.
In the initial stage when there is no HII region, then necessarily
$r_D > r_{ionization}$ and the flow is described simply by the massive molecular
accretion disk. In the second stage, a trapped hypercompact  HII region will develop in
the center of the disk. Because the gas in the disk is denser than the gas elsewhere
around the
star, the molecular accretion disk will not necessarily be fully  ionized, but because
$r_{ionized} < r_G$, the ionized surface of the disk will not be expanding off the
disk. Rather there will be a limited region, contained within the HII region, 
with an ionized accretion flow onto the  disk. 
However, depending on the initial angular momentum in the flow, this
region may be a very small  with respect to 
the extent of the molecular disk. Outside this region, there will be a molecular 
accretion flow onto the disk.
The third stage of evolution is defined by the
condition, $r_{ionized} > r_G$. In the non-spherical  case, because the gas density
around the star is a function of angle off the disk, the extent of ionization, $r_{ionized}$
will also be a function of angle. If the disk is reasonably sharply defined as in the
Ulrich and TSC models, then in the third stage the HII region will  expand
around the disk. In this third stage, because $r_{ionized} > r_G$, the surface
of the disk, with the exception of the small region in the center, 
will be photo-evaporating with an outward flow of ionized gas off
the disk as described in the models for photo-evaporating disks of
Hollenbach et al.~(1994), Johnstone et al.~(1998), Lugo, Lizano \& Garay (2004).

This discussion shows how a consideration of the simple accretion model
of Ulrich (1976) or TSC that is subject to different degrees of ionization can
unify the concepts of the quenched HII region of Walmsley (1995), the 
trapped hypercompact HII regions of Keto (2002b, 2003) and the photo-evaporating disks
of  Hollenbach et al.~(1994) and Johnstone et al.~(1998). In addition the 
three models  are linked in an evolutionary sequence. All three of these models 
will prolong the time scale for accretion onto massive stars, 
and the latter two prolong the lifetimes of the hypercompact HII regions around the
stars. 

In the discussion above, the models assume a smooth accretion flow so that the ionized
zone or HII region has a definite radius. The same theoretical considerations discussed
above also apply to a clumpy
accretion flow if one substitutes for the smooth ionization boundary, the radius
at which the molecular clumps are ionized as they move toward the stars. Since this
radius depends on the density within each clump, there may be a different radius of
ionization for each clump. But on average there will be some characteristic radius 
that differentiates the ionized and molecular zones. In the case of a clumpy
flow, this radius is the radius of ionization. If the gas between the clumps is
ionized as is suggested by the observations of G10.6--0.4, then that gas may move
outward or inward depending on whether it is outside or inside the gravitational
radius, $r_G$. As long as the density remains low, this gas will not have a
strong dynamical influence on the clumpy, molecular accretion flow. 
The structure implied by this discussion of molecular clumps within a low
density ionized flow has been described in some previous studies which we
mention briefly in the section below.

\section{Relevance of the observations to other models of ultracompact HII regions}

The appearance of the photo-ionized arcs seen in figures 2 and 3 is similar to that
expected for
the evaporating or ablating clumps described in models of expanding HII regions proposed by
Dyson, Williams \& Redman (1995), Redman, Williams, \& Dyson (1996), 
and Lizano et al.~(1996). A significant
difference between the HII region proposed in these  models and the one
observed in G10.6--0.4 is that on the scale of
our observations, the ionized gas in G10.6--0.4
is flowing inward as an accretion flow toward the stars and not expanding
outward as in the models. However, if the low
density ionized gas in G10.6--0.4 extends to a scale
beyond the gravitational radius ($\sim$5000 AU), as indicated in lower 
angular resolution observations
(Keto 2002a), while the molecular flow remains clumpy, then
one could imagine that on this larger scale, low density outflowing
ionized gas might stream around molecular clumps similar to 
the description of these models. This picture corresponds more closely to
the model for subsonic outflow in Redman, Williams \& Dyson (1996) rather
than the models for supersonic wind-driven outflows in the other two references.

\section{Considerations for the formation of early type O stars}

As a point of speculation, let us consider the conditions required for early type
O stars, those with masses above about 50 M$_\odot$ where the requirements
are most severe. The introduction discussed
how massive accretion flows with high momentum were required to overcome
the three problems of radiation pressure, thermal pressure, and the short main
sequence life times of massive stars. 

Accretion flows with low angular momentum,
or more precisely low rotational velocities relative to infall velocities
may be required to form early O stars. The reason is that once the accretion flow
settles into a rotationally supported disk, the accretion rate may be reduced if
accretion through the disk is moderated by some type of viscosity as is suspected
for low mass stars. Of course, the stellar radii are so small compared to the
accretion flow that if angular momentum is conserved, the flow must end up
rotationally supported for almost any initial angular momentum no matter how
low. However, the radius of the star may not be the important radius. First, in order
to overcome radiation pressure, the accretion flow must have an 
initial angular momentum that is only low enough that the spin-up through conservation
of angular momentum does not 
significantly reduce the inward momentum
down to the radius of dust destruction. This radius is typically a few
hundred AU as estimated by simple radiative energy balance (e.g.~Scoville \& Kwan 1976).
Secondly, the spin-up itself might be reduced
within the radius of ionization which could be, as in G10.6--0.4, at a distance of
a few thousand AU. Once the flow is ionized,
the gas may couple strongly onto any magnetic fields that may facilitate
the outward transport of
angular momentum and maintain a high rate of inflow to smaller radii. 
One might speculate whether the ionization 
surrounding massive
stars helps rather than hinders the growth of massive
stars by accretion.

\section{Conclusions on star formation by accretion}

The new observations of
G10.6--0.4 show that the detailed structure of the accretion flow exemplifies
several of 
the factors thought to be important in mitigating the three limiting factors,
radiation pressure, thermal pressure, and the short main sequence lifetime of high mass
stars. 

The observations of G10.6--0.4 show that the force deriving from
the inward momentum of the accretion flow, 
$\dot {\rm M} v
\sim 3\times 10^{28}$ dynes, for
$\dot {\rm M} \sim 10^{-3}$ M$_\odot$ yr$^{-1}$ and velocity of $v\sim 4.5$ km s$^{-1}$
is of the magnitude to compete with the
outward force of the radiation pressure, $L/c \sim 1\times 10^{29}$ dynes,
deriving from the
high luminosity
$1.2\times 10^6$ L$_\odot$ of the central stars.
Furthermore, the non-spherical geometry and clumpy structure observed in the accretion
flow should reduce the effective radiation pressure below the simple
estimate above.

The observations of G10.6--0.4 show infall directly through the HII region
despite the thermal pressure of the hot ionized gas. The simple model for 
the accretion flow indicates
that the zone of ionization  of the denser gas is within the gravitational
radius of 5000 AU of the combined mass of the stars. 
Thus the gravitational attraction
is sufficient to counter the thermal pressure of the hot ionized gas,
and the ionized gas as well as
the molecular gas (indicated in these observations by the photoionized arcs) 
will be moving inward as parts of the accretion flow.

The luminosity of G10.6--0.4 indicates the presence of several O stars
(Sollins et al. 2005) that must either have formed in a timescale shorter than their main sequence 
lifetime  or with an accretion rate high enough to delay their main sequence evolution.
The observations indicate a mass accretion rate of about $10^{-3}$ M$_\odot$ yr$^{-1}$
(Keto2002a) on the order of
that required to maintain  O stars on the main sequence as indicated by the timescales
in figure 1.

Thus the new observations of G10.6--0.4 indicate that high mass stars can form by accretion
and indicate how the accretion flow is able to continue even after the stars have 
become massive B and O stars.
Only the most massive 
accretion flows have sufficient momentum to counter the radiation pressure of massive
stars, the high densities to prevent the ionization of the molecular flow at
distances beyond the gravitational radius, $r_G \sim GM/c^2$, and the mass accretion
rate to keep the high mass stars from evolving off the main sequence.

A simple model of accretion on ballistic trajectories (Ulrich 1976, TSC) with the
inclusion of a central ionizing source suggests a unifying paradigm within an
evolutionary sequence for the
individual models of quenched HII regions (Walmsley 1995), trapped HII
regions (Keto 2002b, 2003), and photo-evaporating disks (Hollenbach et al.~1992;
Johnstone et al.~1998).

\section{Acknowledgements}

K.W.~ is funded by a UK PPARC Advanced Fellowship.

\begin{figure}[t]
\includegraphics[width=4.5in,angle=90]{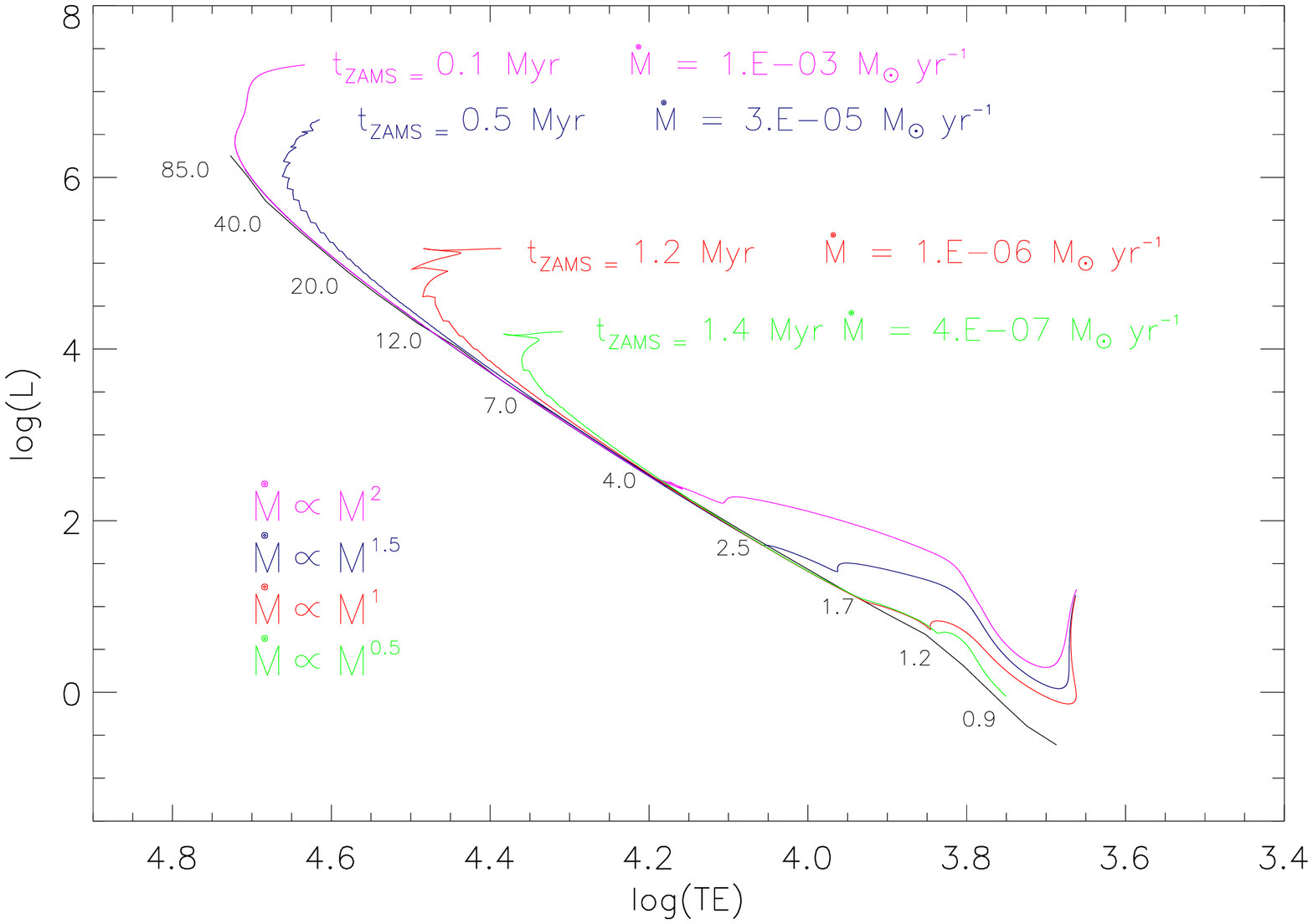}
\caption{HR-diagram with 4 evolutionary tracks for stars
gaining mass by accretion at different rates. 
$t_{\rm ZAMS}$ indicates the time spent evolving up
the ZAMS line from a mass of 4~M$_{\odot}$ star to the mass at which the star
deviates off the ZAMS line.  The accretion rate at the time of deviation is
shown for each track.
}
\label{fig:hr} 
\end{figure}

\begin{figure}[t]
\includegraphics[width=5in,angle=90]{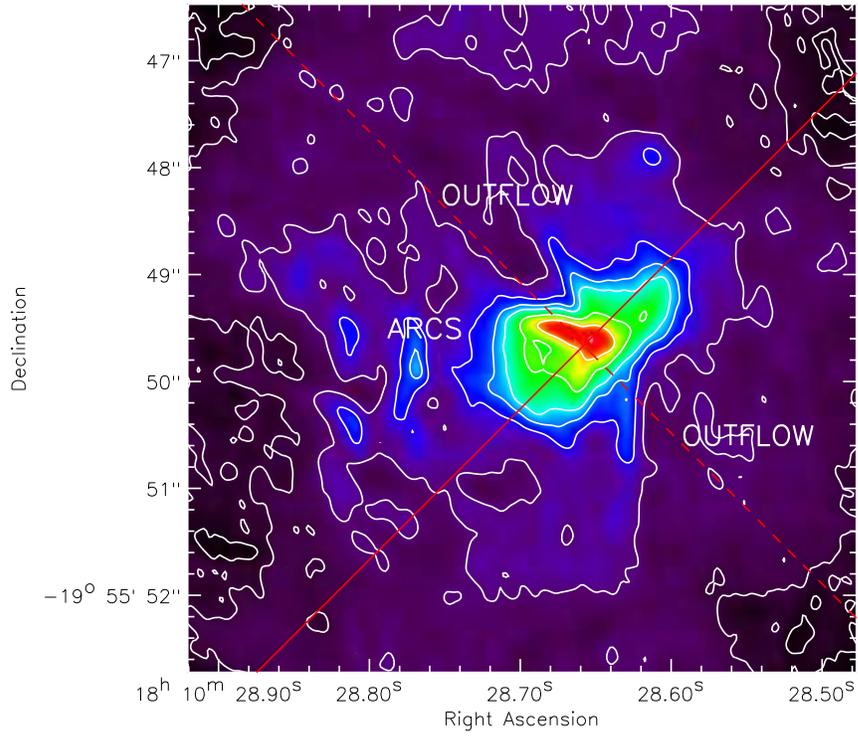}
\caption{
Continuum emission at 1.3 cm.
Solid line and dotted lines: the disk and rotation axes. 
Contour levels are 10\% of the peak flux of 0.04 Jy/beam.
}
\label{fig:continuum} 
\end{figure}

\begin{figure}[t]
\includegraphics[width=5in,angle=90]{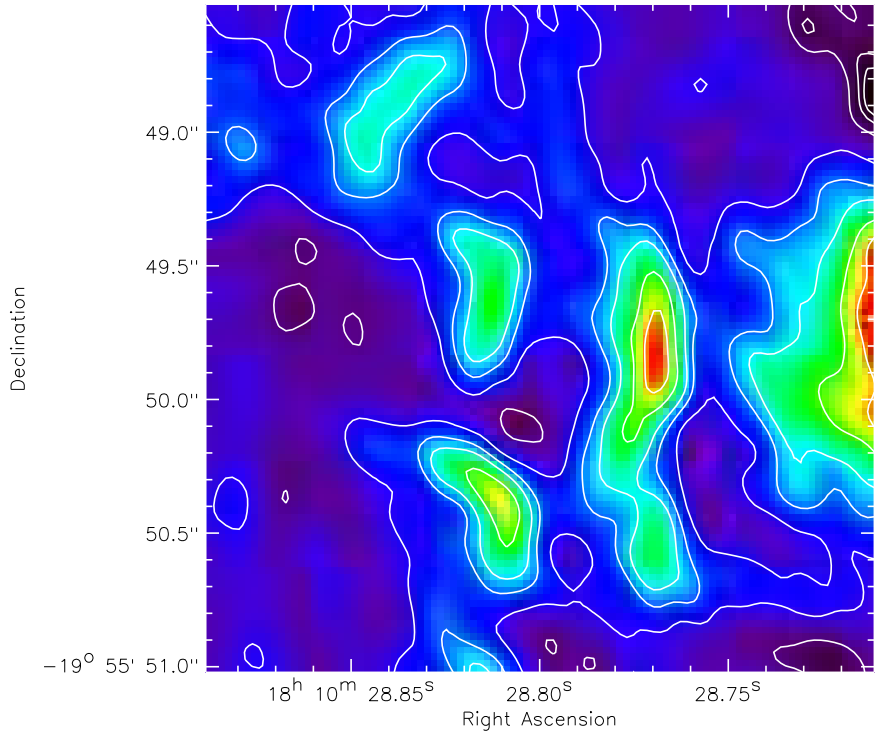}
\caption{
Detail of the arcs seen in 1.3 cm emission. 
Location marked as 'ARCS' on \ref{fig:continuum}.
Contour levels are 10\% of the peak flux of 0.01 Jy/beam.}
\label{fig:arcs}
\end{figure}

\begin{figure}
\vskip -2.5 truein
\includegraphics[width=5in,angle=90]{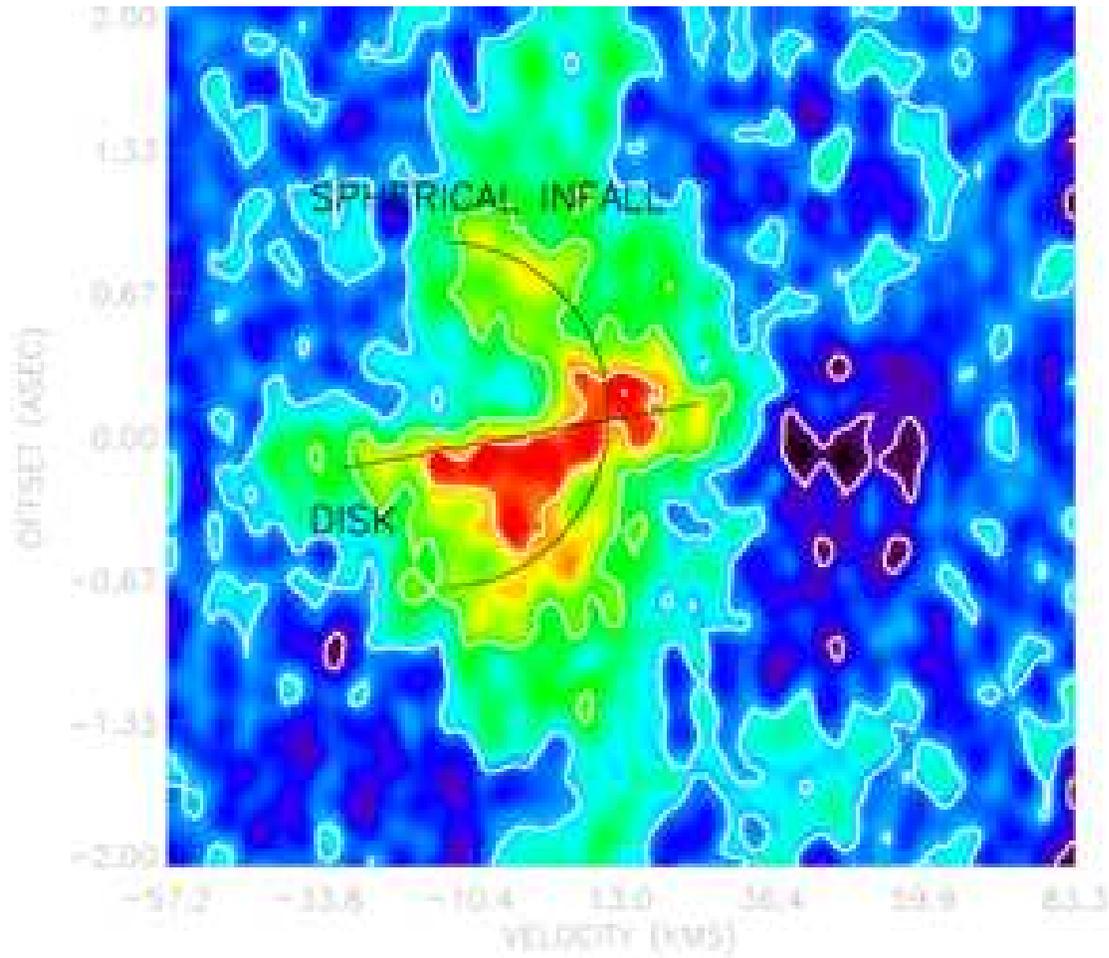}
\caption{
Position versus velocity diagram of the H66$\alpha$ recombination line emission
from the star cluster G10.6--0.4. The position axis lies on a southeast-northwest
line across the long axis of the ionized accretion disk and marked as a solid
line in figure \ref{fig:continuum}.
The infall in the spherical envelope is seen in the backward "C" structure of
the largest scale emission, indicated by the curved black line.
The rotating disk is seen in the brighter, smaller scale
emission at the center of the "C".
The rotation of the disk is seen in the velocity
gradient of
the brightest emission across the disk. The linear gradient of the rotation is
marked by a black line.
The spherical infall is extended on a larger scale along the disk
than perpendicular to the disk (figure \ref{fig:pv_SHORT_disk}) because the scale
and the "C" in the perpendicular direction are flattened due
to the rotation.
The color scale ranges from 0.0 to 0.01 Jy/beam.
The contours are in units of
20\% of the peak emission of 0.01 Jy/beam.
}
\label{fig:pv_LONG_disk}
\end{figure}

\clearpage

\begin{figure}[t]
\vskip -2.5 truein
\includegraphics[width=5in,angle=90]{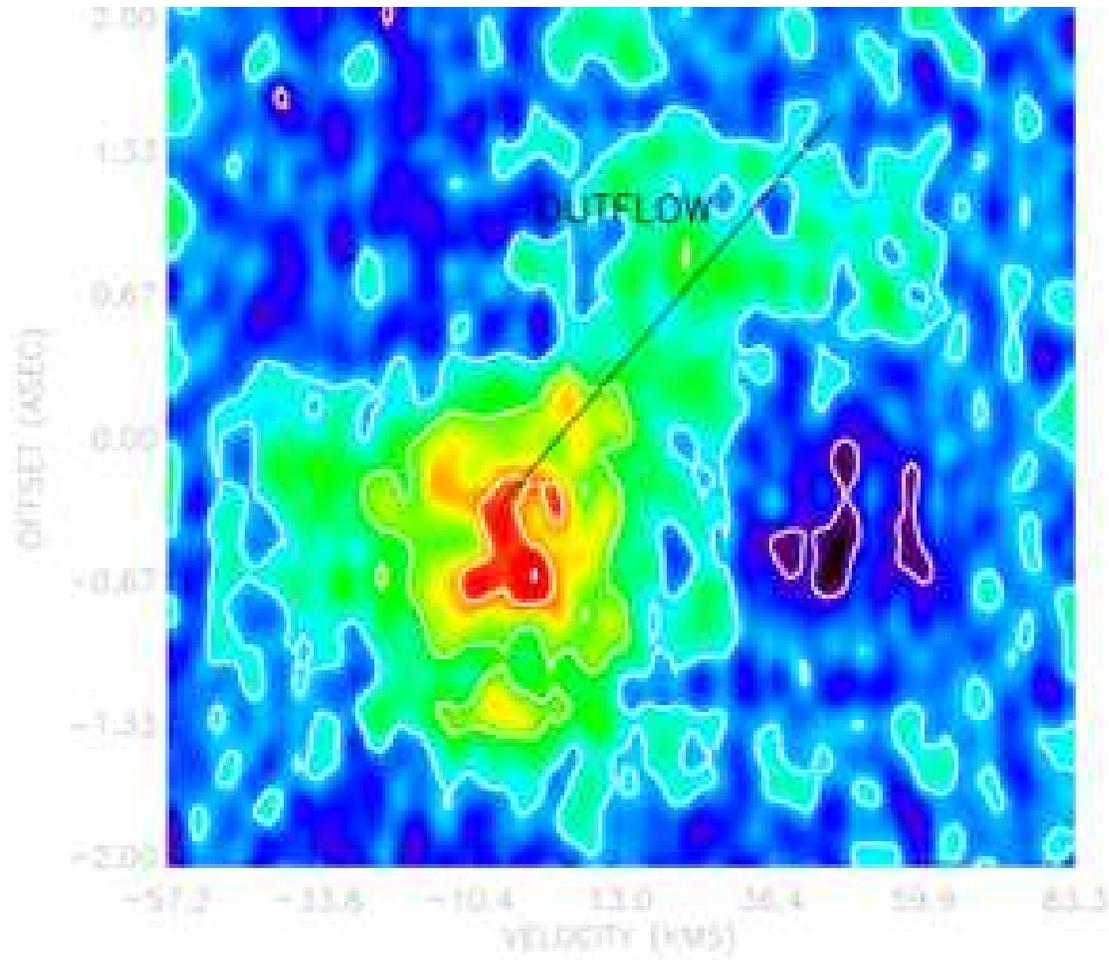}
\caption{
Position versus velocity diagram of the H66$\alpha$ recombination line emission
from the star cluster G10.6--0.4. The position axis,
marked as a dotted line \ref{fig:continuum},
lies on a southwest-northeast
line across the rotation axis and perpendicular to the ionized accretion
disk in figure. 
A high velocity outflow is seen on the northeast side of
the disk extending up to 60 kms$^{-1}$ before falling below the sensitivity limit.
The location of the outflow is indicated by a white line.
The counter outflow on the southwest side of the disk is not seen 
because the emission there is below the sensitivity limit.
The color scale ranges from 0.0 to 0.01 Jy/beam. The contours are in units of
20\% of the peak emission of 0.01 Jy/beam.
}
\label{fig:pv_SHORT_disk}
\end{figure}
\clearpage

\begin{figure}
\includegraphics[width=5in]{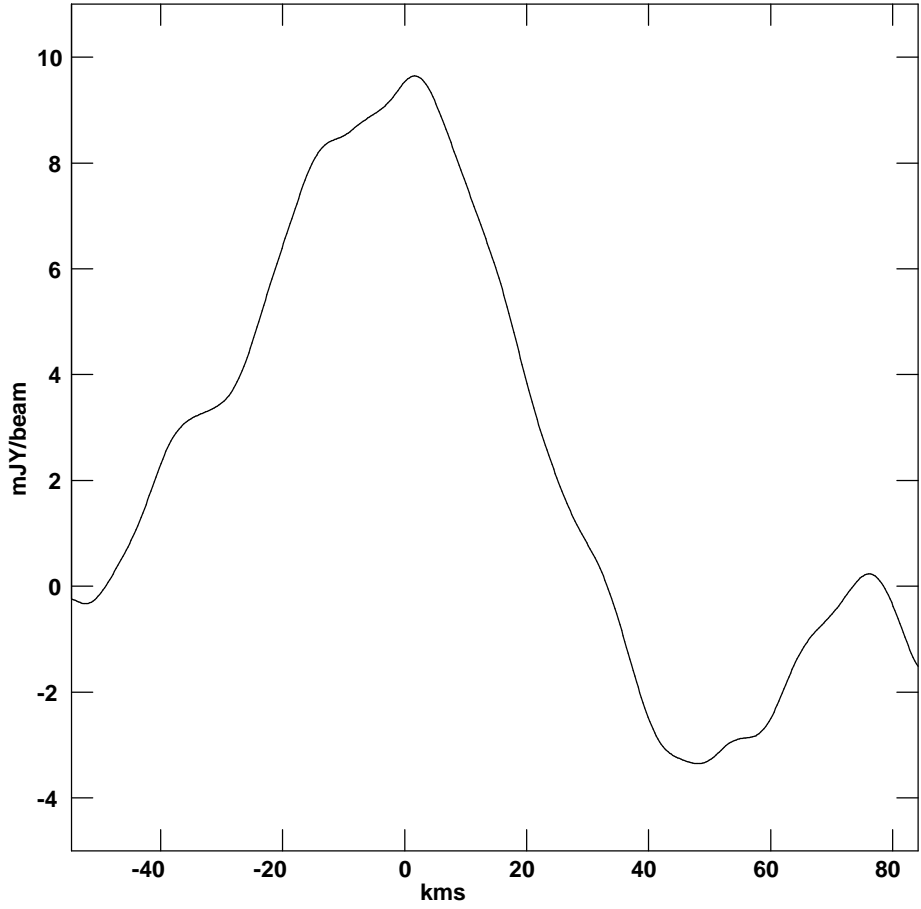}
\caption{
Spectrum of the H66$\alpha$ recombination line at the center of the accretion
flow at the position of the intersection of the solid and dashed lines in
figure \ref{fig:continuum}.
}
\label{fig:spectrum_center}
\end{figure}

\begin{figure}
\includegraphics[width=5in]{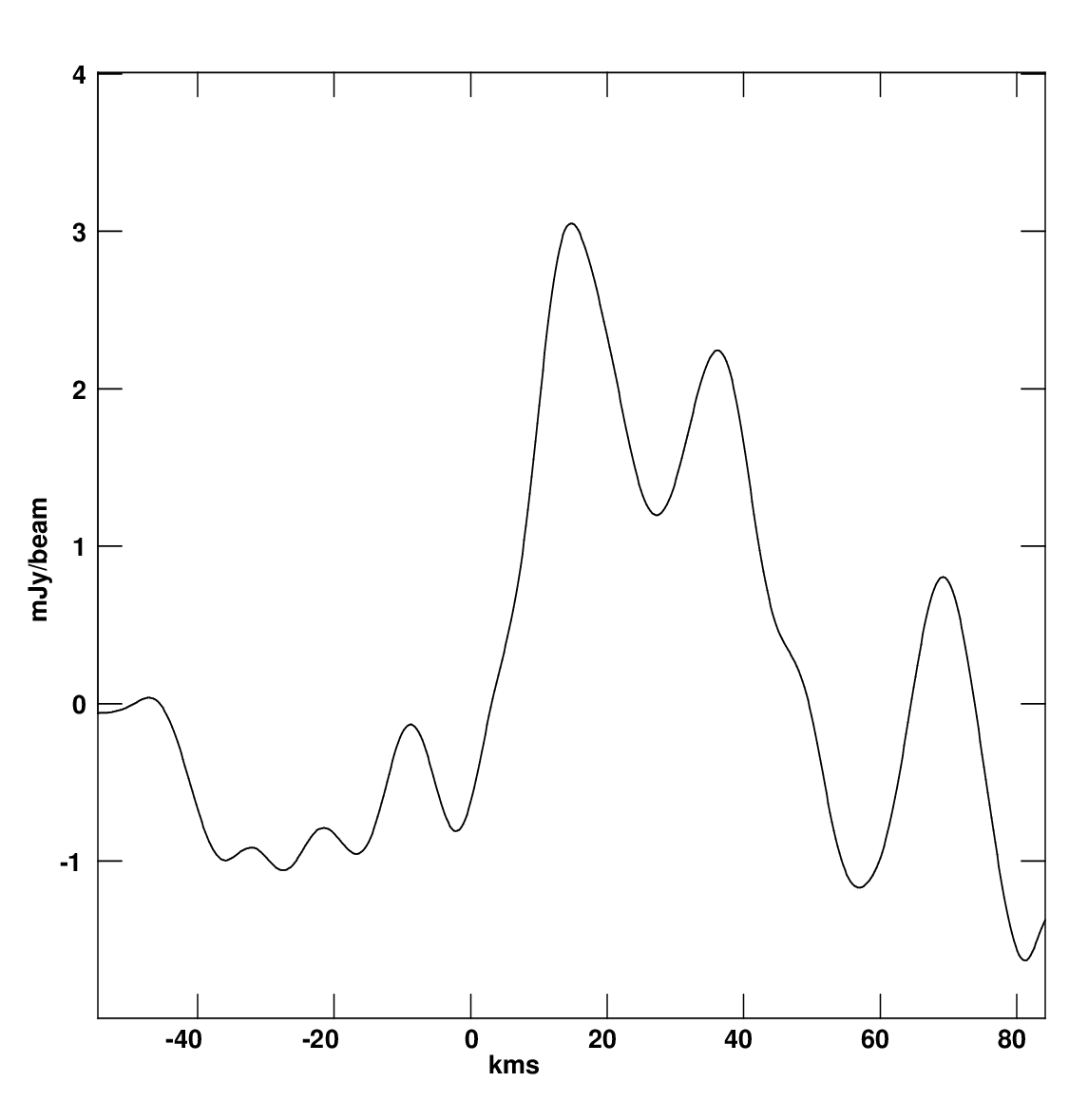}
\caption{
Spectrum of the H66$\alpha$ recombination line in the outflow
at a position along the rotation axis (dashed line in figure \ref{fig:continuum})
1$^{\prime\prime}$ to the southwest
of the intersection of the solid and dashed lines in
figure \ref{fig:continuum}.
}
\label{fig:spectrum_southwest}
\end{figure}

\begin{figure}
\includegraphics[width=5in]{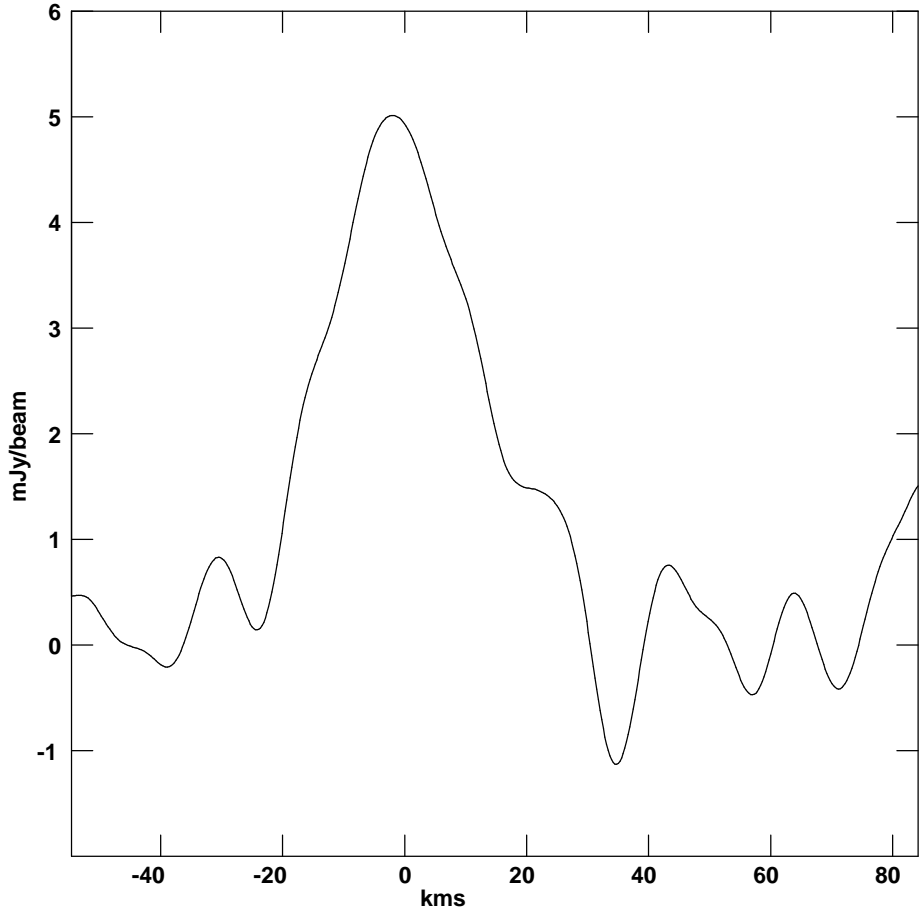}
\caption{
Spectrum of the H66$\alpha$ recombination line in the outflow
at a position along the rotation axis (dashed line in figure \ref{fig:continuum})
1$^{\prime\prime}$ to the northeast
of the intersection of the solid and dashed lines in
figure \ref{fig:continuum}.
}
\label{fig:spectrum_northeast}
\end{figure}

\begin{figure}[t]
\includegraphics[width=5in]{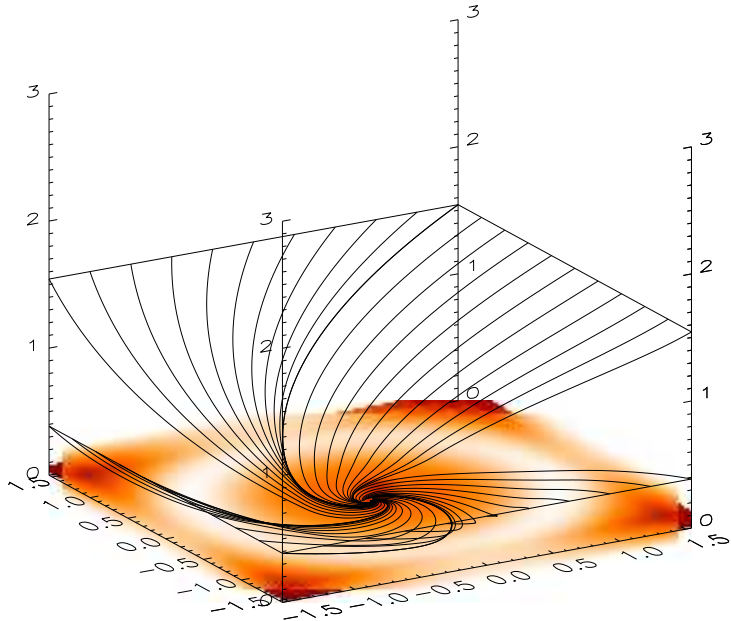}
\caption{
Streamlines of a model of an accretion flow with the gas in ballistic trajectories
around a point mass. The gas starts in a quasi-spherical infall, and owing to
conservation of angular momentum, spins up until a rotationally dominated disk forms
at a non-dimensional radius of unity. The model demonstrates the structure of the
accretion flow onto the cluster G10.6--0.4 that shows quasi-spherical infall on
the larger scales in the molecular gas, and disk accretion on the smaller scales
in the ionized gas.
}
\label{fig:streamlines_3d}
\end{figure}
\clearpage

\begin{figure}[t]
\includegraphics[width=5in]{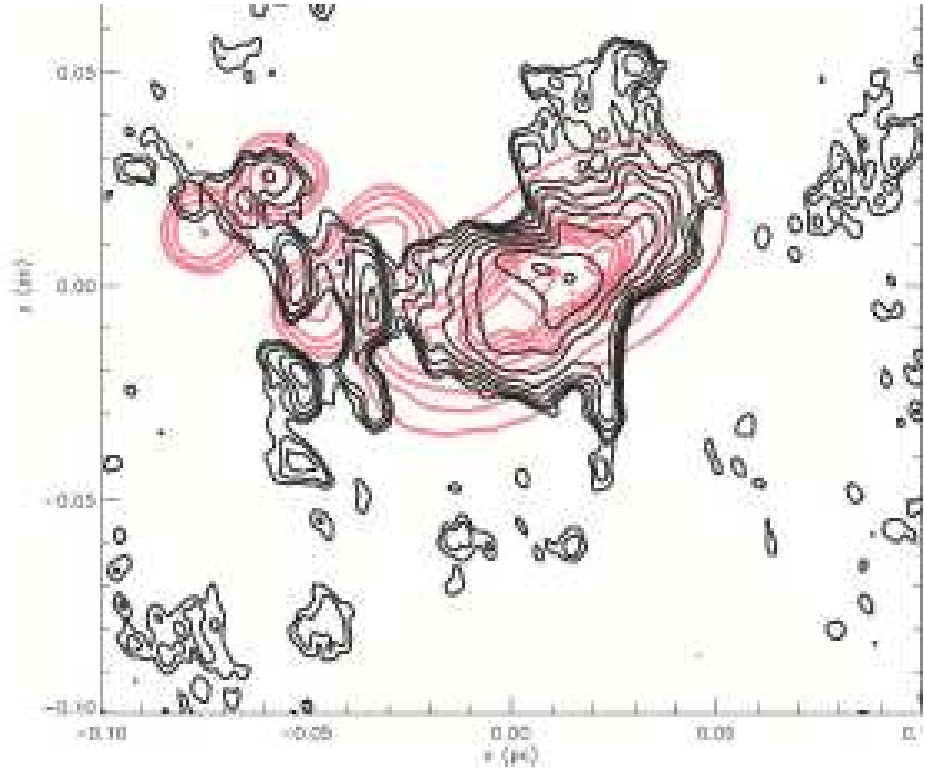}
\caption{
Model of the continuum emission at 1.3 cm from the star cluster G10.6--0.4.
on top of the observed radio continuum.
The model shows an ionized accretion disk and ionized globules
in the clumpy gas around the disk. The model is a
\cite{Terebey1984} accretion disk with a
centrifugal radius of 3500 au, and
an infall rate of $10^{-4}$ M$_\odot$ yr$^{-1}$
onto a 500 M$_\odot$ cluster with additional density fluctuations imposed on
the otherwise smooth structure of the underlying accretion flow. The angular
scale is set for a distance of 6 kpc.
The contour levels in the data start at 1 mJy/beam and increase in
half magnitude levels.
}
\label{fig:datamodel}
\end{figure}
\clearpage

\begin{figure}[t]
\includegraphics[width=5in]{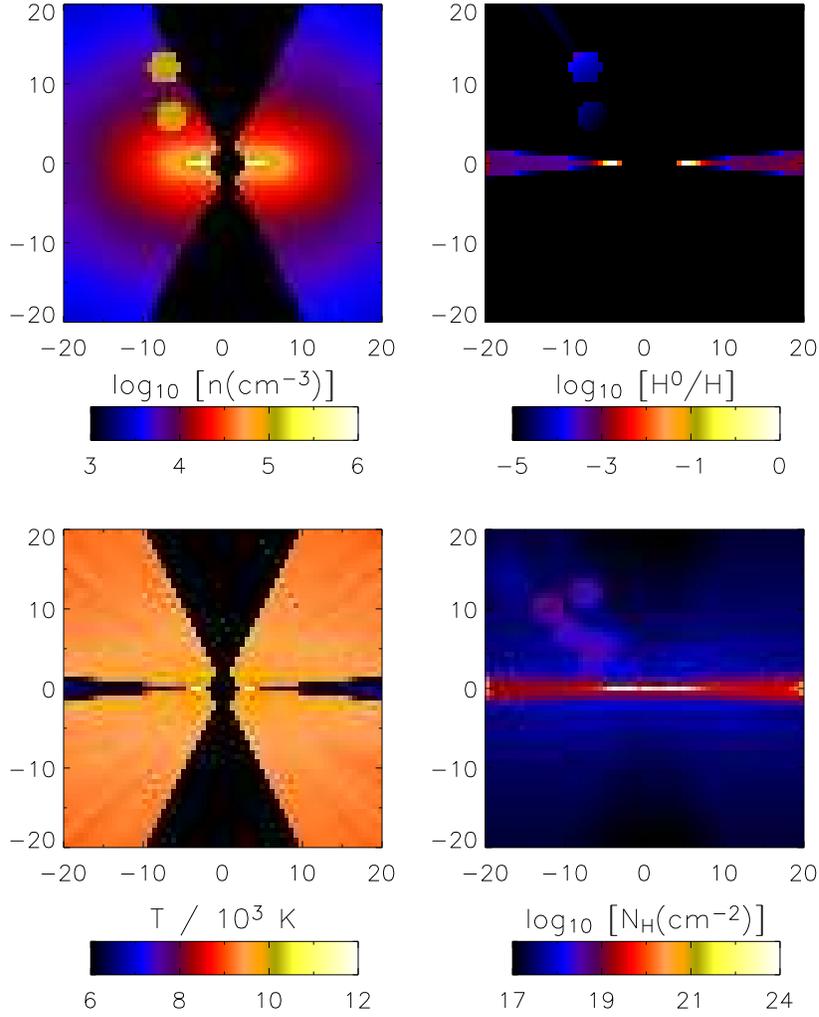}
\caption{
Four slices through the model showing the density,
ionization, temperature, and neutral column density.
The axes are in units
of $10^3$~AU.  The upper left panel
shows the density, the evacuated bipolar cavity,
and two of the six clumps included in our model.  The lower left
panel shows the gas temperature. The upper right panel shows the
ionization fraction.
The lower right panel shows the column density of neutral gas.
}
\label{fig:ionstruc}
\end{figure}
\clearpage

\end{document}